\documentclass[a4paper]{article}
\usepackage[margin=25mm]{geometry}
\usepackage{amsmath}
\usepackage{amsfonts}
\usepackage{amssymb}
\usepackage{graphicx}
\usepackage{verbatim}
\usepackage{bm}
\usepackage{mathtools}
\usepackage{amssymb}
\usepackage{tkz-euclide}
\usepackage{tensor}
\usepackage{hyperref}
\usepackage[sorting=none, style=nature, backend=biber]{biblatex}
\usepackage{biblatex} 
\providecommand{\keywords}[1]
{
  \small	
  \textbf{\textit{Keywords---}} #1
}

\title{Topological constraints on general relativistic galaxies: Exploring 
novel conical singularity networks}
\author{Marco Galoppo$^{1,*}$\\ \\
        \small $^1$School of Physical \& Chemical Sciences, University of Canterbury, \\
        \small Private Bag 4800, Christchurch 8041, New Zealand \\ \\
        \small $^*$marco.galoppo@pg.canterbury.ac.nz
}
\date{\today} 

\begin{document}
\pagenumbering{arabic}
\maketitle

\begin{abstract}

The van Stockum-Bonner class of spacetimes can be interpreted as fully general relativistic models for rigidly rotating disc galaxies. Frame-dragging effects in these geometries demand a recalibration of the dark matter content relative to models based on Newtonian gravity. We investigate the previously overlooked topological structure of these spacetimes, in relation to the viability of fully general relativistic galaxy toy models. We discuss the appropriate boundary conditions for these solutions to model disc galaxies. For this class of spacetimes, we show the existence of a network of quasi-regular singularities along the rotation axis of the galaxies. The existence of such novel conical defect structures further restricts the physical viability of the van Stockum-Bonner class. Unwinding these issues is key to avoiding pathologies in future fully general relativistic modelling of alternative to dark matter.

\end{abstract}


\maketitle 

\section{\label{sec:Intro}Introduction}

The description of galactic dynamics is the subject of ongoing debate in astrophysics. Indeed, a purely Newtonian description of galaxies is in irreconcilable conflict with the observed flat rotation curves \cite{NConc1,NConc3,NConc4,NConc5,NConc6}. To resolve this, many approaches have been developed including MOdified Newtonian Dynamics (MOND) \cite{MOND1,MOND2,MOND3,MOND4,MOND5,MOND6,MOND7,MOND8,MOND9}, MOdified Gravity (MOG) theories \cite{MOG} and the Dark Matter (DM) hypothesis \cite{DM}. Of these, the latter, namely the assumption of the existence of a significant non-baryonic component of matter, is arguably the most widely accepted. 
\\
\\
The DM hypothesis is indeed one of the foundations of the standard $\Lambda$ Cold Dark Matter ($\Lambda$CDM) cosmological model \cite{Planck}. It has been highly successful in interpreting a plethora of different astrophysical observations, such as: rotation curves of disc galaxies \cite{NConc1,NConc3,NConc4,NConc5,NConc6}; velocity distribution of galaxies in Galaxy Clusters (GCs) \cite{NConc2,GCs}; thermodynamic properties of X-ray emitting gas in GCs \cite{DMok1}; gravitational lensing produced by GC mass distributions \cite{DMok2}; features of the two Bullet Clusters  \cite{DMok3,DMok4,DMok5} ; the growth of cosmic structures from inhomogeneities in the matter density content of the early universe 
\cite{DMok6,DMok7}.
Nonetheless, despite such remarkable success, the results of the many experiments aimed at the direct detection of DM particles are, to date, inconclusive \cite{directdm0,directdm1,directdm2,directdm3,directdm4,directdm5,directdm6}.  In addition, several observations have challenged the validity of the $\Lambda$CDM model \cite{LCDM1,LCDM2,LCDM3,LCDM4}. In particular, if we only focus on galaxies, several recent independent observations appear to conflict with the standard DM paradigm \cite{Gal1,Gal2,Gal3,Gal4,Gal5,Gal6,Gal7,Gal8,Fluffy1,Fluffy2,Fluffy3, JWST1,JWST2,JWST3,JWST4,JWST5,JWST6,JWST7}. 
\\
\\
Given the challenges to the DM hypothesis, as well as to MOND and MOG theories \cite{Crit2}, a new approach is gaining traction, namely the use of exact solutions of Einstein's equations to model cosmological structures. Thus far, this approach has been applied only to model features of individual galaxies \cite{Tieu1,Tieu2,Tieu3,BG,Crosta,Beordo2023,SergioVittorio,Federico1,Astesiano2,Galoppo}. Ultimately, it is important to extend this approach to the dynamics of galaxies within clusters. However, important topological issues have been overlooked in refs.\ \cite{Tieu1,Tieu2,Tieu3,BG,Crosta}, which need to be addressed in order to consistently model complex superposition of such structures.   
\\
\\
The new approach rests on the highly nonlinear nature of General Relativity (GR). The nonlinearities of GR introduce novel ingredients as compared to both Newtonian physics and special relativity. In such theories the transition between particles and effective fluid description is well understood in terms of coarse-graining and averaging. In GR any nongravitational binding energy is supplemented by quasilocal sources of gravitational energy: spacetime itself carries dynamical energy and angular momentum. 
\\
\\
Neglecting these quasilocal terms in galactic modelling is usually justified by invoking the weak field limit  -- the typical velocities of stars and gas are nonrelativisitc, $\beta = v/c \approx 10^{-3}$. Nonetheless, the weak field presupposes a background, conventionally Minkowski spacetime, $g_{\mu\nu} = \eta_{\mu\nu}+h_{\mu\nu}$, $|h_{\mu\nu}|\ll 1$, and the process of calibrating one Minkwoski background from one system to another in which is embedded can be highly nontrivial
\cite{David1,David2,David3,David4}. This occurs already in the transition from the few body systems in which GR is directly tested (stars, black holes \dots) to many body systems (star clusters, galaxies, GCs \dots) \cite{David5}. Thus, the typical weak field limit will fail when na\"{\i}vely applied to large scale systems such as galaxies. Furthermore,  Ciotti and collaborators \cite{Ciotti1,Ciotti2,Ciotti3} proved that even a perturbative implementation of the standard gravitomagnetic limit is not feasible as a solution for galactic modelling without DM. Consequently, if the DM phenomenon is to be resolved within a GR framework, it will be in a nonlinear/nonperturbative regime.
\\
\\
Balasin and Grumiller presented a full GR galaxy model to address the DM phenomenon \cite{BG}. Crosta and coworkers \cite{Crosta,Beordo2023} showed that the Balasin-Grumiller model (BG) fits the Milky Way (MW) rotation curve reconstructed from the GAIA satellite's kinematic data without any need for DM \footnote{We note that in ref.\ \cite{BG} the authors claimed a reduction of only 30\% of DM. However, this can be traced to a na\"{\i}ve comparison of the densities of BG and classical models.}. Moreover, the BG model was shown to be preferred to MOND or DM-driven galactic dynamics, on account of a reduced set of free parameters and similar goodness of fit to the rotation curve. Nevertheless, despite its clear successes, the BG solution fails both to model the galaxy bulge \cite{BG,Crosta,Costa} and match gravitational lensing observations \cite{Galoppo}. In particular: (i) average stellar motion in the galactic bulge is not circular; and (ii) close encounters of stars are frequent enough to invalidate treating matter as a pressureless fluid. Furthermore, the rigid rotation assumption \cite{BG} leads to unphysical time delay differences which rule out inferences about strong lensing \cite{Galoppo}.
\\
\\
To bypass the deficiencies of rigid rotation \cite{Stephani,Islam}, Cacciatori and collaborators \cite{SergioVittorio,Federico1} have investigated the full class of stationary, axisymmetric dust solutions of Einstein equations with boundary conditions appropriate for disc galaxies. Naturally, this begs the question of what ``appropriate'' boundary conditions are. To better appreciate this, in the present paper we investigate the presence and physical interpretation of topological defects and conical singularities for the entire rigidly rotating van Stockum--Bonner (vSB) class\footnote{This includes the BG \cite{BG} and Cooperstock-Tieu \cite{Tieu1,Tieu2,Tieu3} models.} \cite{vSBclass1,vSBclass2,vSBclass3,vSB,vSBclass4}. In our work, we draw on well-established mathematical results \cite{EllisSchimdt,Conical1,BezzeraBook,Conical2,Anderson} which have played a role in interpreting the physical nature of topological defects in various settings, i.e. formation in early universe phase-transitions \cite{Phase1,Phase2}, gravitational lensing \cite{TopDefectLensing1,TopDefectLensing2} and shifts of atomic spectra \cite{TopDefectAtom1,TopDefectAtom2}.
\\
\\
The structure of this paper is as follows: in section \ref{sec:CS} we introduce the concept of quasi-regular singularities and conical singularities; in section \ref{sec:Gm} we define the vSB models, we discuss the observers through which we read their physics, and we specialise to BG; in section \ref{sec:ResGen} we discuss appropriate boundary conditions for galaxy models, prove the existence of nonisolated quasi-regular singularities in the vSB class and describe the resulting topological features; section \ref{sec:Conc} is dedicated to a brief overview of the results and the discussion of future perspectives. 

\section{\label{sec:CS}Quasi-regular singularities}
A point $q$ of a spacetime $(M,\boldsymbol{g})$ is defined as a quasi-regular singularity if it is the end point of an incomplete geodesic $\gamma(\lambda)$, when $\lambda$ is a generalised affine parameter, and it is not a curvature singularity \cite{Conical1,EllisSchimdt}. Namely, the curvature components $R_{\mu\nu\rho\sigma}$ measured in an orthonormal frame are continuous in $q$. The critical property of all quasi-regular singularities is their undetectability from local considerations. Indeed, local quantities are well-behaved when evaluated in any open set containing a quasi-regular singularity. Instead, their presence is imprinted on the global topological structure of the spacetime. 
\\
\\
We are interested in the quasi-regular singularities named conical singularities, such as the one identified by the tip of a cone. To understand their nature, let us consider Minkowski spacetime in cylindrical coordinates
\begin{equation}
    ds^2 = -dt^2 + dr^2 +r^2d\phi^2 + dz^2, \label{Minkwoski}
\end{equation}
where $t,z \in (-\infty,\infty)$, $r \in [0,+\infty)$ and $\phi \in [0,2\pi]$. To obtain from this spacetime a conical singularity, we may proceed by identifying points related by the translation \cite{Conical1,EllisSchimdt}
\begin{equation}
    \phi = \phi + \alpha, \label{translation}
\end{equation}
where $\alpha \neq 2\pi$. This results in a new spacetime with the same metric as \eqref{Minkwoski} but for which $t,z \in (-\infty,\infty)$, $r \in [0,+\infty)$ and $\phi \in [0,\alpha]$. The points on the z-axis in this new spacetime are conical singularities. To gauge their singular nature, we compute the circumference-to-radius ratio for any circle drawn around the z-axis on a 2-surface \{$t = const, z = const$\} in the limit $r \longrightarrow 0$. Doing so gives us a ratio exactly equal to $\alpha$. instead of the Euclidean $2\pi$. These singularities result focusing (attractive) if $\alpha < 2\pi$ and defocusing (repulsive) if $\alpha > 2\pi$. 
\\
\\
We point out that the presence of conical singularities in a spacetime can be immediately inferred whenever its line element can be cast in the form
\begin{equation}
    ds^2 = -dt^2 + dr^2 + b^2 r^2 d\phi^2 + dz^2.
    \label{cosmicstring}
\end{equation}
Indeed, \eqref{cosmicstring} is equivalent to \eqref{Minkwoski} but with $0 < \phi < 2\pi b$, as can be seen by applying the change of coordinates $\phi ' = b\phi$. The identification of conical singularities is thus generally achieved by writing the respective line element in a form equivalent to \eqref{cosmicstring}, as in the case of cosmic strings \cite{BezzeraBook, Anderson}. More complex procedures exist to identify conical singularities, e.g., calculating the holonomy group of the manifold for the suspected singular points \cite{Conical2}. However, these have proven unnecessary in our current work.

\section{\label{sec:Gm}Galaxy Models}
The vSB spacetimes are a subclass of the galaxy models class which models disc galaxy dynamics using a stationary, axisymmetric metric expressed in standard cylindrical coordinates  
\begin{align}
    ds^2 = & g_{tt}(r,z)dt^2 + 2g_{t\phi}(r,z)dt d\phi \nonumber \\  &+ g_{\phi\phi}(r,z)d\phi^2 + e^{\mu(r,z)}\left(dr^2+dz^2\right),
    \label{metric1}
\end{align} 
where we use the convention c = 1 and the metric is coupled to
a dust energy-momentum tensor of the form
\begin{equation}
    T_{\mu\nu} = \rho(r,z)u_\mu u_\nu.
\end{equation}
The coupling has been worked out in \cite{Islam,Stephani}. Thus, we have
\begin{equation}
    u^\mu \partial_\mu = \sqrt{-H}\left(\partial_t+\Omega \partial_\phi\right),
\end{equation}
\begin{align}
    &g_{tt}=\frac{\left(H-\eta\Omega \right)^2-r^2\Omega^2}{H} \label{4_g00} ,\\
    &g_{t\phi}=\frac{r^2-\eta^2}{H}\ \Omega+\eta \label{4_g03} ,\\
    &g_{\phi\phi}=\frac{\eta^2-r^2}{H}, \label{metric2}
\end{align}
\begin{align}
    &\mu_{,r}=\frac{1}{2r}\left[g_{tt,r}g_{\phi\phi,r}-g_{tt,z}g_{\phi\phi,z}  -\left(g_{t\phi,r}\right)^2+\left(g_{t\phi,z}\right)^2\right] , \label{mur}\\
    &\mu_{,z}=\frac{1}{2r}\left[g_{tt,z}g_{\phi\phi,r}-g_{tt,r}g_{\phi\phi,z}-2g_{t\phi,z}g_{t\phi,r}\right], \label{muz}
\end{align}
\begin{equation}
    8\pi G \rho = \frac{\left(\eta_{,r}^2+\eta_{,z}^2\right)\left[\eta^2r^{-2}\left(2-\ell\eta\right)^2-r^2\ell^2 \right]}{4\eta^2 e^\mu}, \label{etaHdensity}
\end{equation}
where $\eta$ is a function of $r$ and $z$, $H$ is an arbitrary negative function of $\eta$, $\ell = H'/H$ is the logarithmic derivative of $H$ and $\Omega$ is defined as
\begin{equation}
    \Omega  \coloneqq \frac{1}{2}\int \frac{H'}{\eta}d\eta.
\end{equation}
The parameter $\Omega (r,z)$ describes the angular velocity of the dust referred to the coordinates in use 
\begin{equation}
    \Omega  = \frac{d\phi}{dt}
\end{equation}
The function $\eta(r,z)$ can be implicitly retrieved through 
\begin{equation}
    \mathcal{F} = 2\eta + r^2\int \ell(\eta) \left( \frac{1}{\eta}-\eta\right) d\eta,
    \label{f1}
\end{equation}
where, as a consequence of Einstein's equations, $\mathcal{F}$ satisfies the harmonic equation
\begin{equation}
    \mathcal{F}_{,rr}- \frac{1}{r}\mathcal{F}_{,r} +\mathcal{F}_{,zz} = 0.
    \label{f2}
\end{equation}
Equation \eqref{f2} corresponds to the differential equation for $\eta(r,z)$
\begin{equation}
    \left(\eta_{,rr}-\frac{1}{r}\eta_{,r}+\eta_{,zz}\right)\left(2-\eta\ell(\eta)\right) + \left(\eta_{,r}^2-\eta_{,z}^2\right)\left[\ell'(\eta)\left(\frac{r^2}{\eta}+\eta\right)-\ell(\eta)\left(1+\frac{r^2}{\eta^2}\right)\right] + r^2\frac{\ell(\eta)}{\eta}\left(\eta_{,rr}+-\frac{3}{r}\eta_{,r}+\eta_{,zz}\right) = 0.
    \label{etaH_equation_eta}
\end{equation}
\eqref{etaH_equation_eta} uniquely determines $\eta(r,z)$ once $H(\eta)$ and $\eta(r,0)$ are arbitrarily assigned. From here on out, we refer to this class of galaxy models as the $(\eta,H)$ class.
\subsection{The ZAMO observers}
A physical interpretation of the field $\eta(r,z)$ is achieved once we choose the appropriate class of observers for the galaxy. In the case of stationary, axisymmetric metrics, there exists a natural class of observers in terms of which to read the physics of the system: the Zero Angular Momentum Observers (ZAMO) 
 \cite{Zamo1,Zamo2}. These are defined by the tetrad
\begin{align}
    & \boldsymbol{e^0} = \frac{r}{\sqrt{g_{\phi\phi}}} dt,\\
    & \boldsymbol{e^1} = e^{\mu/2}dr,\\
    & \boldsymbol{e^2} = e^{\mu/2}dz,\\
    & \boldsymbol{e^3} = \sqrt{g_{\phi\phi}} \left(d\phi -\chi dt\right). 
\end{align}
We define the velocity of the dust in the galaxy as measured by the reference frame formed by the ZAMO, $v(r,z)$, through
\begin{equation}
     -e^0_\mu u^\mu \eqqcolon \frac{1}{\sqrt{1-v^2}} 
    \label{ZAMO1}
\end{equation}
where $u^\mu$ is the four-velocity of the dust. On the other hand, we also have
\begin{equation}
    -e^0_\mu u^\mu = \frac{\sqrt{-H}r}{\sqrt{g_{\phi\phi}}} = \frac{1}{\sqrt{1-\left(\eta/r\right)^2}}.
    \label{ZAMO2}
\end{equation}
Thus, we can identify $\eta(r,z)$ as 
\begin{equation}
    \eta(r,z) = r  v(r,z).
    \label{ETABABY}
\end{equation}
Therefore, the field $\eta(r,z)$ is henceforth understood as the product of the velocity field of the dust, measured in the reference frame built by ZAMO, times the radial coordinate. 
\\
\\
Naturally, the question arises of whether the velocity profile measured by ZAMO can be interpreted as the one measured in astronomical observations. This view, held true in refs.\ \cite{Tieu1,Tieu2,Tieu3,BG,Crosta,Beordo2023} has been challenged by Costa and collaborators \cite{Costa}. However, we believe the original point of view to be correct, at least when implemented in the analysis of the rotation curve obtained by GAIA for the MW (see ref.\ \cite{CrostaZamo1,CrostaZamo2}). As such, we will consider ZAMO to be the fiducial observers henceforth.

\subsection{van Stockum-Bonner Galaxies}
To obtain the equations defining the sVB class we must make the mutually inclusive choices 
\begin{align}
    & H(\eta) = -1 , \label{cond_yup_1} \\
    & \Omega(\eta) = 0, \label{cond_yup_2}
\end{align}
which imply that vSB galaxies undergo rigid rotation, a clear drawback of this class of models. From \eqref{4_g00},\eqref{4_g03},\eqref{metric2},\eqref{cond_yup_1} and \eqref{cond_yup_2}, we have the metric terms as
\begin{align}
    & g_{tt} = -1, \label{gtt}\\
    & g_{t\phi} = \eta, \label{gtphi}\\ 
    & g_{\phi\phi} = r^2-\eta^2 \label{gphiphi}.
\end{align}
So that
\begin{equation}
    ds^2 = -dt^2 + 2\eta(r,z) dt d\phi + (r^2-\eta^2)d\phi^2 + e^{\mu(r,z)}(dr^2+dz^2). \label{vSBLineElement}
\end{equation}
From \eqref{mur}, \eqref{muz}, \eqref{gtt}, \eqref{gtphi} and \eqref{gphiphi} we get
\begin{align}
    & \mu_{,r} = -\frac{1}{2r}\left(\eta_{,r}^2 - \eta_{,z}^2\right), \label{murvSB} \\
    & \mu_{,z} = -\frac{1}{2r}\eta_{,r}\eta_{,z} \label{muzvSB}.
\end{align}
Through the use of \eqref{cond_yup_1} and \eqref{cond_yup_2}, \eqref{etaH_equation_eta} reduces to
\begin{equation}
    \eta_{,rr} - \frac{\eta_{,r}}{r} + \eta_{,zz} = 0.\label{etavSB}
\end{equation}
\eqref{murvSB}. \eqref{murvSB} and \eqref{etavSB}  completely solve for the vSB spacetime class. The general solution to \eqref{etavSB} is given by \footnote{To see this, it is sufficient to substitute $\eta(r,z) = rf(r,z)$ in \eqref{etavSB}. The equation reduces to $ - \triangle f + \frac{1}{r^2} = 0$,
where $\triangle$ is the laplacian in cylindrical coordinates. The resulting equation is the static Schrödinger equation for $E = 0$, for a particle with mass $m = \hbar^2/2$ in a central potential of the form $V(r) = 1/r^2$. It is a well-known result \cite{QM1,QM2,QM3} that any solution of this equation can be written as an integral over the eigenvalues of its separable solutions.} 
\begin{equation}
    \eta(r,z) = \int_0^{+\infty} \left[A(\lambda)\cos(\lambda z) + B(\lambda) \sin(\lambda z)\right]\lambda r K_{1}(r\lambda) d\lambda + \eta_{c} = \hat{\eta}(r,z) + \eta_{c}, 
    \label{generalEta}
\end{equation}
where $K_{1}(r\lambda)$ is the MacDonald function of the first order, $A(\lambda)$ and $B(\lambda)$ are, respectively, the spectral densities for the even and odd modes of the solution and $\eta_{c}$ is the constant of integration. Finally, from \eqref{etaHdensity}, \eqref{cond_yup_1} and \eqref{cond_yup_2}, the dust density is given by
\begin{equation}
    8\pi G \rho = \frac{\eta_{,r}^2+ \eta_{,z}^2}{r^2e^\mu} \label{densitivytvSB}
\end{equation}
\subsection{Balasin-Grumiller Galaxy}
The BG model is a specific solution of the vSB class obtained by choosing \cite{BG}
\begin{align}
    & A(\lambda) = \frac{2}{\pi}\int_0^{+\infty} C(x) \cos{(\lambda x)},\label{1choice}\\
    & B(\lambda) = 0,\label{2choice}\\
    & \eta_{c} = -\int_0^{+\infty} A(\lambda)d\lambda\label{3choice},
\end{align}
where \eqref{2choice} corresponds to choosing a galaxy symmetrical with respect to the equatorial plane and $C(x)$ is given by 
\begin{align}
    C(x) = &V_0 \left[(x-r_0)(\theta(x-r_0)-\theta(x-R)\right] \nonumber \\
    & + V_0(R-r_0)\theta(x-R), \label{weirdBG}
\end{align}
where $V_0$ is the asymptotic velocity measured by ZAMO, $R$ is the galaxy radius and $r_0$ is the bulge radius. \eqref{1choice}, \eqref{2choice}, \eqref{3choice} and \eqref{weirdBG} lead to the analytic expression for $\eta(r,z)$
\begin{align}
    \eta(r,z) = & \frac{V_0}{2}\sum_{\pm} \left(\sqrt{(z\pm r_0)^2 + r^2}-\sqrt{(z\pm R)^2 + r^2}\right) \nonumber \\
   & +  V_0(R-r_0).
    \label{BGETA}
\end{align}
The other relevant quantities of the system are obtained through the equations previously discussed for the entire class of models.

\section{\label{sec:ResGen}Conical Singularities in van Stockum-Bonner Galaxies}
\begin{figure*}[htb!]
\centering 
\includegraphics[width=0.65\textwidth]{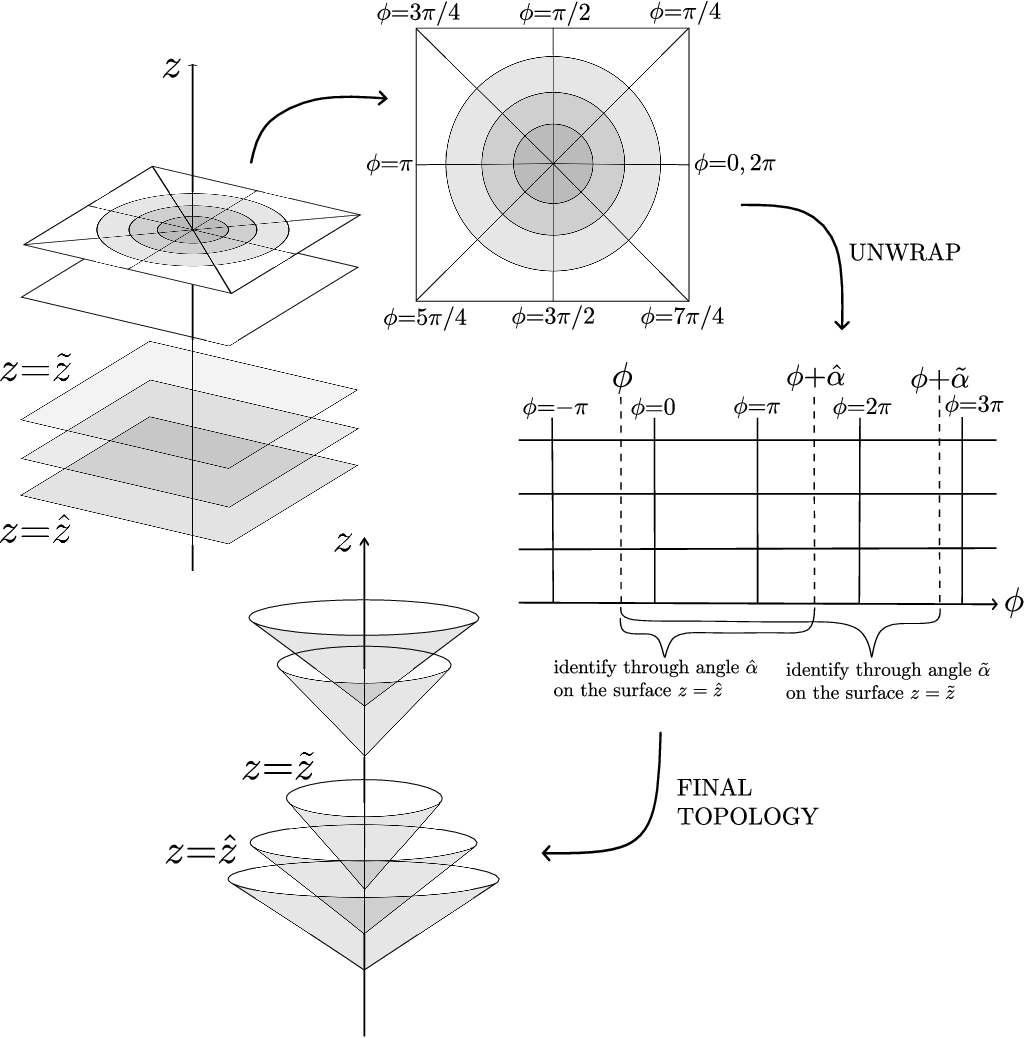}
\caption{Topology around the z-axis of an vSB galaxy. The slicing and identification procedure for different {$z = const$} is shown.}
\label{complex_topology}
\end{figure*}

To investigate the topological structure of the vSB class, we start by considering the asymptotic line element at radial infinity 
\begin{equation}
    ds^2 \simeq -dt^2 + 2\eta_{c}dt d\phi + r^2d\phi^2 + e^{\mu(+\infty,z)}\left(dr^2+dz^2\right), \label{radial_infinity}
\end{equation}
where $\mu(+\infty,z) = \lim_{r \longrightarrow +\infty}\mu(r,z)$ and we have used that $\lim_{r \longrightarrow +\infty}\hat{\eta}(r,z) = 0$ (see \eqref{generalEta}). As we are interested in using vSB spacetimes as galaxy models, we notice that the persistence of the off-diagonal term at spatial infinity is undesirable. Full GR galaxy models must be smoothly matched at large distances from the matter bulk with a void Kerr-like solution of Einstein's equations. Therefore, the term $g_{t\phi}(r,z)$ must present an asymptotic behaviour of the type $\propto A/r$, where $A$ is a given constant. However, this is not true if $\eta_{c}$ is non-null. In particular, the clocks of the fiducial observers placed at radial infinity will be desynchronised\footnote{This is precisely what happens in the BG model, as shown in \cite{Galoppo}.}, hindering their very role as reliable asymptotic inertial observers. Therefore, we have found the first boundary condition for the vSB galaxy models
\begin{equation}
    \eta_{c} = 0. \label{first_constraint}
\end{equation}
We can further specialise the writing of $\mu(+\infty,z)$. From \eqref{murvSB} and \eqref{muzvSB} we have
\begin{equation}
    \mu(r,z) = - \int_0^z \frac{\eta_r(r,z')\eta_{z'}(r,z')}{2r}dz' + F(r) + \mu_c,
\end{equation}
where $F(r)$ must be so that \eqref{murvSB} is satisfied and $\mu_c$ is the integration constant. Since from \eqref{etavSB} we get $ \lim_{r \longrightarrow +\infty}\eta_{,r}(r,z) = \lim_{r \longrightarrow +\infty}\eta_{,z}(r,z) = 0$, we have
\begin{equation}
    \mu(+\infty,z) = \lim_{r \longrightarrow +\infty} F(r) + \mu_c. \label{mu_infinity}
\end{equation}
Therefore, \eqref{mu_infinity} shows that $\mu(+\infty,z) = \mu_{\infty}$. Moreover, we have the freedom to choose
\begin{equation}
    \mu_c = - \lim_{r \longrightarrow +\infty} F(r).\label{second_constraint}
\end{equation}
Imposing \eqref{second_constraint} and \eqref{first_constraint} is equivalent to requiring an asymptotic Minkowskian structure at infinity. This is a fair boundary condition for would-be galaxy models, given their necessary matching to void Kerr-like solutions. However, the vanishing of the Riemannian tensor, $R_{\mu\nu\rho\sigma}$, furnish twenty independent relationships and, yet, only ten curvature components, $R_{\mu\nu}$, enter into the laws of the gravitational field. Therefore, even the less strict boundary condition of local flatness would still be a physically sound requirement for these spacetimes. Henceforth, we impose \eqref{second_constraint} in conjunction with \eqref{first_constraint} to facilitate calculations. Nonetheless, we hold the view that a simply locally flat spacetime, as the one generated by a cosmic string (see \eqref{cosmicstring}), would still be physically acceptable. 
\\
\\
As we have discussed the appropriate boundary conditions at spatial infinity, we can shift our focus to investigate the presence of quasi-regular singularities. To gauge the presence of a conical singularity, we must consider the behaviour of $g_{\phi\phi}(r,z)$ and $g_{rr}(r,z)$ next to the rotation axis at a fixed value of $z$. This is equivalent to studying, respectively, the limiting behaviour of $\eta(r,z)$ and $\mu(r,z)$. To the leading order, the MacDonald function $K_{1}(x)$ reads \cite{MacDonald}
\begin{align}
    K_1(x) \simeq \frac{1}{x} + o(x\log(x)). \label{McDonaldSmall}
\end{align}
Therefore, we get
\begin{equation}
    \eta(r,z)_{\big{|}r \ll 1} = \int_0^\infty \left[A(\lambda)\cos{(\lambda  z)} + B(\lambda)\sin{(\lambda z)}\right] d\lambda .  \label{eta_close_axis}
\end{equation}
For global galaxy models \eqref{eta_close_axis} implies
\begin{equation}
    \int_0^\infty \left[A(\lambda)\cos{(\lambda z)} + B(\lambda)\sin{(\lambda z)}\right] d\lambda = 0 ~ \forall ~ z . \label{CONSTRAINT1}
\end{equation}
Indeed, if \eqref{CONSTRAINT1} was not satisfied, $g_{\phi\phi}(r,z)$ would necessarily become negative close to the galaxy centre. Though it is true that most galaxies possess a central supermassive black hole, we are investigating the possibility of globally modelling a galaxy using vSB models. Hence, we believe \eqref{CONSTRAINT1} to be a reasonable requirement on the metric function\footnote{This condition, even though reasonable, it is not entirely necessary. Indeed, it can not be realised, saved for a singular plane, for any galaxy model possessing reflection symmetry with respect to the equatorial plane (for which $B(\lambda) = 0$), such as the BG solution. Thus, the subclass of symmetrical sVB galaxies models is forced to be considered as producing viable solutions only in a well-defined domain which excludes the bulge of the galaxy. However, this does not disqualify these models as effective outside the bulge of a disc galaxy. Indeed, even the Newtonian description of gravity is haunted by the presence of singularities, i.e. in the Newtonian potential for a point particle.  Nonetheless, it is clearly a perfectly valid physical description on its scales of applicability.}. Furthermore, by applying the same reasoning,  we require
\begin{align}
    & \lim_{r \longrightarrow 0} \mu_{,r}(r,z) \neq \pm \infty \label{mu_constraint_1} \\
    & \lim_{r \longrightarrow 0} \mu_{,z}(r,z) \neq \pm \infty \label{mu_constraint_2}
\end{align}
\eqref{mu_constraint_1} and \eqref{mu_constraint_2} are equivalent to
\begin{equation}
    \lim_{r \longrightarrow 0} \mu(r,z) = f(z)  ~ s.t. ~ |f(z)| < +\infty. \label{CONSTRAINT2}
\end{equation}
To study the condition \eqref{CONSTRAINT2}, we must investigate the behaviour of $\eta_{,r}(r,z)$ and $\eta_{,z}(r,z)$ for small $r$. From \eqref{etavSB} we get
\begin{align}
    &\eta_{,r}(r,z)_{|r \ll 1} = o(r\log(r)) \Rightarrow \lim_{r \longrightarrow 0} \eta_{,r}(r,z) = 0, \label{lim1} \\
    &\lim_{r \longrightarrow 0} \eta_{,z}(r,z) = \int_0^{+\infty} \left[B(\lambda) \cos(\lambda z)-A(\lambda)\sin(\lambda z)\right]\lambda d\lambda, \label{lim2} 
\end{align}
where we have used the expansion of $K_0(x)$ \cite{MacDonald} in \eqref{lim1}
and \eqref{McDonaldSmall} to obtain \eqref{lim2}. Thus, \eqref{murvSB} and \eqref{CONSTRAINT2} give a new condition on the spectral densities 
\begin{equation}
    \int_0^{+\infty} \left[B(\lambda) \cos(\lambda z)-A(\lambda)\sin(\lambda z)\right]\lambda d\lambda = 0. \label{CONSTRAINT3}
\end{equation}
We notice that to satisfy \eqref{mu_constraint_2}, given \eqref{muzvSB} and \eqref{lim1}, it is sufficient for the spectral densities to meet the condition \eqref{CONSTRAINT3}. We can prove that if \eqref{CONSTRAINT1} is satisfied, so is \eqref{mu_constraint_1}. Let us define
\begin{equation}
    T(\lambda) = \theta(\lambda)\left[A(\lambda)+ i B(\lambda)\right]. \label{T}
\end{equation}
\eqref{T} allows us to write \eqref{CONSTRAINT1} as
\begin{equation}
    \mathcal{F}\left(T(\lambda)\right)(z) = - \left[\mathcal{F}\left(T(\lambda)\right)(z)\right]^*, \label{Fourier1}
\end{equation}
where $\mathcal{F}$ indicates the Fuorier-Plancherel transform and $^*$ denotes complex conjugation. By using the basic properties of the Fuorier-Plancherel transform, \eqref{CONSTRAINT3} becomes
\begin{equation}
    \frac{d}{dz}\left[\mathcal{F}\left(T(\lambda)\right)(z)\right] = -  \frac{d}{dz}\left[\left(\mathcal{F}\left(T(\lambda)\right)(z)\right)^*\right], \label{Fourier2}
\end{equation}
which follows directly from \eqref{Fourier1}. Therefore, a vSB model satisfying \eqref{first_constraint}, \eqref{second_constraint} and \eqref{CONSTRAINT1} represents a global full GR galaxy model as it produces an asymptotically globally Minkowskian spacetime at spatial infinity and the metric functions are well-behaved over the whole coordinate domain. However, these models are found to still harbour quasi-regular singularities.  Let looks at the asymptotic form of the line element near the rotation axis
\begin{equation}
    ds^2 \simeq -dt^2 + r^2 d\phi^2 + e^{f(z)}\left(dr^2 +dz^2)\right), \label{close1}
\end{equation}
where we have considered \eqref{CONSTRAINT2}. Let us specialise to any 2D surface \{$t = const, z = const = \hat{z}$\}. The line element reads
\begin{equation}
    ds_{2D}^2 = r^2 d\phi^2 + e^{f(\hat{z})}dr^2, \label{close2}
\end{equation}
On the chosen 2D surface, we define $r = e^{f(\hat{z})/2}\Tilde{r}$ so that the line element in \eqref{close2} becomes
\begin{equation}
    ds_{2D}^2 = e^{-f(\hat{z})}\Tilde{r}^2 d\phi^2 + d\Tilde{r}^2 = b^2(\hat{z})\Tilde{r}^2 d\phi^2 + d\Tilde{r}^2. \label{closeFINAL}
\end{equation}
\eqref{closeFINAL} is exactly equivalent to the 2D surface line element which signals the presence of conical singularities (see \eqref{cosmicstring}). Moreover, given \eqref{muzvSB}, $f(z)$ could be null only for models cylindrically symmetric or invariant under radial translations, clearly unphysical conditions. Thus, any vSB global galaxy model defines a spacetime with a highly nontrivial topology. Indeed, each 2D spacetime slice of the type \{$t = cost, z = cost$\} harbours a conical singularity in $r = 0$.  However, unlike more well known cases, i.e. cosmic strings, the conical structure is more complex, and it is entirely defined by the function $f(z)$. We can interpret the topological structure of these vSB spacetimes \footnote{Notice that the proof of the presence of quasi-regular singularities still holds even if \eqref{second_constraint} is not satisfied.} as obtained by slicing the spacetime near the rotation axis along 2D surfaces of the type {$z = const$} and folding each one in a cone with a varying angle of identification $\alpha$. Fig. \ref{complex_topology} shows a tentative visualisation of this topology. 
\\
\\
We must stress that not all the points along the rotation axis will be quasi-regular singularities. Indeed, for asymptotically flat vSB spacetimes, a mixture of quasi-regular and curvature singularities should be expected \cite{vSB}. In particular, for the BG solution it was shown that two limited disconnected region of the z-axis harbour curvature singularities \cite{Costa}. 
\\
\\
Naturally, the presence of these singularities begs the question about their ultimate cause in the vSB class. A possible explanation may come from the unphysical condition of rigid rotation engrained in these spacetimes. If so, these problematic features could be cured by considering differentially rotating models in the larger $(\eta, H)$ class. However, the singularities might also be the result of the absence of pressure in the dust or even of enforcing axial symmetry on the spacetimes. All the aforementioned causes are worthy of consideration and should be thoroughly researched. 
\\
\\
As it stands, the presence of a nontrivial structure of quasi-regular singularities in vSB global galaxy solutions is in flat contrast with the very possibility of describing globally a galaxy with such models. Therefore, any vSB solution with reasonable asymptotic properties can only be considered as a viable galaxy model only for a limited portion of the galaxy. In particular, these models will fail in describing the bulge of the galaxy. However, we must stress that the limitation of applicability of vSB solutions in no way rules them out as effective galaxy models over their region of applicability. Any such model which, under the correct choice of reference frame, were to correctly account for current astronomical observations should be regarded as a functioning effective full GR modelling of galactic dynamics. Indeed, any physical model de facto posses a domain of applicability, beyond which it breaks down. Nonetheless, this does not preclude its employment in explaining physical observations in its domain of validity. The vSB class must be discarded as a viable choice for global, full GR galaxy models but it should still be considered as possibly producing domain-limited effective full GR galaxy models.
\\
\\
Finally, given the presence of conical singularities, the proper definition of physical coordinates must be put into question. Indeed, in GR coordinates are a priori devoid of a physical meaning. They acquire one only when a measurement procedure is defined. The impact of conical singularities on the definition of the angle coordinate for physical observers inside a galaxy must be correctly addressed when using full GR models. If we specialise to sVB metrics, we notice that the conical singularity on the equatorial plane can always be negated by a proper choice of $\mu_c$. We would argue that it is precisely this choice that a physical observer would take. However, such a choice, which validates the angular coordinate's common physical meaning, fixes a degree of freedom of the model. Therefore, any physical quantity calculated will be directly impacted by this choice -- i.e. density and rotation curve calculations should be carried out only once the coordinate choice has been defined in a physically sound way.

\section{\label{sec:Conc}Conclusions and Perspectives}

In 1973 John Archibald Wheeler famously summarised General Relativity as: \textit{Space tells matter how to move, 
matter tells space how to curve} \cite{Gravitation}. This historical quote directly points to the geometrical nature of GR. Indeed, GR describes space-time as a four-dimensional pseudo-Riemannian manifold whose local geometry is everywhere defined by its matter-energy content through Einstein's equations, but it does not prescribe its global topological structure. Nonetheless, topological questions play a crucial role in our understanding of the Universe \cite{HawkingEllis,Singular0,Singular1,Singular2,Singular3,Singular4,Singular5,Singular6,Topology1,Topology2,Orientability}, since topology can \textit{limit} possible matter content.
\\
\\
Here we have used spacetime topology in this  limiting fashion. In particular, even at the level of modelling galaxies in full GR, topological consideration must be thoroughly studied. They can restrict the scale of viability of a model or rule it out entirely. 
\\
\\
In this paper, we showed that the van Stockum-Bonner class does not contain physically viable full GR global galaxy models. Well-behaved solutions, including any larger class of asymptotic flat geometries, were shown in sec. \ref{sec:ResGen} to be studded with quasi-regular singularities along the rotation axis. These conical singularities generate a highly nontrivial topological structure (see Fig. \ref{complex_topology}) which starkly contrast with their interpretation as global galaxy models. 
\\
\\
The vSB class was shown to contain only effective full GR galaxy models, whose domain of physical validity must necessarily be restricted outside the galactic bulge. Therefore, any serious search for a global, full GR galaxy model should focus on a larger class of models -- possibly the $(\eta,H)$ class.
\\
\\
The $(\eta,H)$ class investigated by Cacciatori \textit{et al} \cite{SergioVittorio,Federico1} which generalises the vSB class may prove to be a fruitful line of enquiry. By dropping the rigidity of the dust and allowing for differential rotation, the class of models is physically realistic. However, the inclusion of an effective pressure might still be necessary to fully model the galactic bulge and avoid singularities. As such, the introduction of pressure in full GR galaxy modelling still remains a high priority of the field. Nonetheless, we have shown the potential for topological considerations in model selection by limiting the class of viable global full GR galaxy models. We plan to make full use of these considerations in future work.

\section*{Acknowledgments}
We thank David Wiltshire, Sergio Cacciatori and Chris Harvey-Hawes for useful discussion and Morag Hills for providing the plot in section \ref{sec:ResGen}.

\printbibliography 

@PREAMBLE{
 "\providecommand{\noopsort}[1]{}" 
 # "\providecommand{\singleletter}[1]{#1}%" 
}

@ARTICLE{NConc1,
   author       = "F. Zwicky",
   title        = "Die {R}otverschiebung von extragalaktischen {N}ebeln",
   journal      = "Helvetica Physica Acta",
   volume       = "6",
   pages        = "110",
   year         = "1933",
}

@ARTICLE{NConc2,
  author = "F. Zwicky",
  title = "On the {M}asses of {N}ebulae and of {C}lusters of {N}ebulae",
  journal = "Astrophysical Journal",
  volume = "86",
  pages = "217",
  year = "1937",
}

@ARTICLE{NConc3,
  author = "J.H. Oort",
  title = "{S}ome {P}roblems {C}oncerning the {S}tructure and {D}ynamics of the {G}alactic {S}ystem and the {E}lliptical {N}ebulae {NGC} 3115 and 4494",
  journal = "Astrophysical Journal",
  volume = "91",
  pages = "273",
  year = "1940",
}

@ARTICLE{NConc4,
  author = "Rubin V. C. and W. K. Ford Jr",
  title = "Rotation of the {A}ndromeda {N}ebula from a {S}pectroscopic {S}urvey of {E}mission {R}egions",
  journal = "Astrophysical Journal",
  volume = "159",
  pages = "379",
  year = "1970",
}

@ARTICLE{NConc5,
  author = "Rubin V. C. and W. K. Ford Jr",
  title = "Extended rotation curves of high-luminosity spiral galaxies. {IV} – {S}ystematic dynamical properties, {SA} through {SC}",
  journal = "The Astrophysical Journal Letters",
  volume = "225",
  pages = "L107-L111",
  year = "1978",
}

@ARTICLE{NConc6,
  author = "Y. Sofue and V. C. Rubin",
  title = "Rotation curves of spiral galaxies",
  journal = "Ann.Rev.Astron.Astrophys.",
  volume = "39",
  pages = "137",
  year = "2001",
}

@ARTICLE{MOND1,
  author = "B. Famaey and S. S. McGaugh ",
  title = "{M}odified {N}ewtonian {D}ynamics ({MOND}): {O}bservational {P}henomenology and {R}elativistic {E}xtensions",
  journal = "Living Reviews in Relativity",
  volume = "15",
  pages = "10",
  year = "2012",
}

@ARTICLE{MOG,
  author = "S. Nojiri, S.D. Odintsov and V.K. Oikonomou",
  title = "Modified gravity theories on a nutshell: {I}nflation, bounce and late-time evolution",
  journal = "Physics Report",
  volume = "692",
  pages = "1",
  year = "2017",
}

@ARTICLE{DM,
  author = "Bing-Lin Young",
  title = "A survey of dark matter and related topics in cosmology",
  journal = "Frontiers of Physics",
  volume = "12",
  year = "2016",
}

@ARTICLE{Crit2,
  author = "D. C. Rodrigues and V. Marra and A. del Popolo and Z. Davari",
  title = "Absence of a fundamental acceleration scale in galaxies",
  journal = "Nature Astronomy",
  volume = "2",
  pages = "668",
  year = "2018",
}

@MISC{Tieu1,
  author = "F.I. Cooperstock and S. Tieu",
  title = "General relativity resolves galactic rotation without exotic dark matter",
  howpublished = "e-print arXiv:astro-ph/0507619", 
   month        = "", 
   year         = "2005", 
   note         = "",
}

@ARTICLE{Tieu2,
  author = "F.I. Cooperstock and S. Tieu",
  title = "Galactic dynamics via general relativity: a compilation and new developments",
  journal = "Int. J. Mod. Phys.",
  volume = "22",
  pages = "2293",
  year = "2007",
}

@ARTICLE{Tieu3,
  author = "J. Carrick and Fred I. Cooperstock",
  title = "General relativistic dynamics applied to
  the rotation curves of galaxies",
  journal = "Astrophys. Space Sci",
  volume = "337",
  pages = "321",
  year = "2012",
}

@ARTICLE{BG,
  author = "H. Balasin and D. Grumiller",
  title = "Non-{N}ewtonian behavior in weak field
general relativity for extended rotating sources",
  journal = "Int. J. Mod. Phys. D",
  volume = "17",
  pages = "475",
  year = "2008",
}

@ARTICLE{Crosta,
  author = "M. Crosta and M. Giammaria and M.G. Lattanzi and E. Poggio",
  title = "{CDM} and geometry-driven {M}ilky {W}ay rotation curve models with {G}aia {DR2}",
  journal = "Mon. Not. R. Astron. Soc.",
  volume = "496",
  pages = "2107",
  year = "2020",
}

@ARTICLE{SergioVittorio,
  author = "D. Astesiano and S. L. Cacciatori and V. Gorini and F. Re",
  title = "Towards a full general relativistic approach to galaxies",
  journal = "Eur. Phys. J. C ",
  volume = "82",
  pages = "554",
  year = "2022",
}

@MISC{Federico1,
  author = "F. Re and D. Astesiano and S. L. Cacciatori and M. Dotti and V. Gorini and F. Haardt",
  title = "Re-weighting dark matter in disc galaxies: a new general relativistic observational test",
  howpublished = "e-print arXiv:astro-ph:2204.05143", 
   year         = "2022", 
   note         = "",
}

@ARTICLE{Astesiano2,
  author = "D. Astesiano and M. L. Ruggiero",
  title = "Galactic dark matter effects from purely geometrical aspects of general relativity",
  journal = "Phys. Rev. D",
  volume = "106",
  year = "2022",
}

@MISC{Galoppo,
  author = "M. Galoppo and S.L. Cacciatori and V. Gorini and M. Mazza",
  title = "{Equatorial Lensing in the Balasin-Grumiller Galaxy Model}",
  howpublished = "e-print arXiv:gr-qc:2212.10290", 
   month        = "December", 
   year         = "2022", 
   note         = "",
}

@article{Ciotti1,
  author = "L. Ciotti",
  title = "On the {R}otation {C}urve of {D}isk {G}alaxies in {G}eneral {R}elativity",
  journal = "The Astrophysical Journal",
  volume = "936",
   year = "2022", 
}

@MISC{Ciotti2,
  author = "Kostas Glampedakis and David Ian Jones",
  title = "{Pitfalls in applying gravitomagnetism to galactic rotation curve modelling}",
  howpublished = "e-print arXiv:astro-ph:2303.16679", 
   month        = "March", 
   year         = "2023", 
   note         = "",
}

@MISC{Ciotti3,
  author = "Kostas Glampedakis and David Ian Jones",
  title = "{Does gravitational confinement sustain flat galactic rotation curves without dark matter?}",
  howpublished = "e-print arXiv:astro-ph:2303.11094", 
   month        = "March", 
   year         = "2023", 
   note         = "",
}

@BOOK{Stephani,
   author       = {H. Stephani and D. Kramer and M. MacCallum and  C. Hoenselaers and E. Herlt},
   year         = 2003,
   title        = {Exact {S}olutions of {E}instein’s {F}ield {E}quations},
   publisher    = {Cambridge University Press}
}

@BOOK{Islam,
   author       = {J. N. Islam},
   year         = 2009,
   title        = {Rotating {F}ields in {G}eneral {R}elativity},
   publisher    = {Cambridge University Press}
}

@ARTICLE{Zamo1,
    author = "J. M. Bardeen",
  title = "A variational principle for rotating stars in {G}eneral {R}elativity",
  journal = "The Astrophysical Journal",
  volume = "162",
  pages = "71",
  year = "1970",
}

@ARTICLE{Zamo2,
  author = "J. M. Bardeen and W. H. Press and S. A. Teukolsky",
  title = "Rotating {B}lack {H}oles: {L}ocally {N}onrotating {F}rames, {E}nergy {E}xtraction, and {S}calar {S}ynchrotron {R}adiation",
  journal = "The Astrophysical Journal",
  volume = "178",
  pages = "347",
  year = "1972",
}

@ARTICLE{Fluffy1,
  author = "P. van Dokkum and others",
  title = "A galaxy lacking dark matter",
  journal = "Nature",
  volume = "555",
  pages = "629",
  year = "2018",
}

@ARTICLE{Fluffy2,
  author = "S. Trujillo-Gomez and J. M. D. Kruijssen and M. Reina-Campos",
  title = "The emergence of dark matter-deficient ultra-diffuse galaxies driven by scatter in the stellar mass–halo mass relation and feedback from globular clusters",
  journal = "Mon. Not. R. Astron. Soc.",
  volume = "510",
  pages = "3356",
  year = "2021",
}

@article{Fluffy3,
year = {2022},
volume = {936},
author = {Demao Kong and Manoj Kaplinghat and Hai-Bo Yu and Filippo Fraternali and Pavel E. Mancera Piña},
title = {The Odd Dark Matter Halos of Isolated Gas-rich Ultradiffuse Galaxies},
journal = {The Astrophysical Journal},
}

@ARTICLE{DMok1,
  author = "Allen and others",
  title = "{Cosmological Parameters from Observations of Galaxy Clusters}",
  journal = "ARAA",
  volume = "409",
  year = "2011",
}

@INPROCEEDINGS{DMok2,
       author = {{Taylor}, A.},
        title = "{Gravitational lens magnification and galaxy cluster masses}",
    booktitle = "{Evolution of Large Scale Structure: From Recombination to Garching}",
         year = 1999,
       editor = {{Banday}, A.~J. and {Sheth}, R.~K. and {da Costa}, L.~N.},
        month = jan,
        pages = {226},
}

@article{DMok3,
year = {2006},
volume = {648},
number = {2},
pages = {L109},
author = {Douglas Clowe and Maruša Bradač and Anthony H. Gonzalez and Maxim Markevitch and Scott W. Randall and Christine Jones and Dennis Zaritsky},
title = "{A Direct Empirical Proof of the Existence of Dark Matter*}",
journal = {The Astrophysical Journal},
}

@article{DMok4,
year = {2008},
volume = {687},
number = {2},
pages = {959},
author = {Maruša Bradač and Steven W. Allen and Tommaso Treu and Harald Ebeling and Richard Massey and R. Glenn Morris and Anja von der Linden and Douglas Applegate},
title = "{Revealing the Properties of Dark Matter in the Merging Cluster MACS J0025.4–1222*}",
journal = {The Astrophysical Journal},
}

@article{DMok5,
year = {2012},
volume = {747},
pages = {96},
author = { M.J. Jee and A. Mahdavi and H. Hoekstra and others},
title = "{A Study of the Dark Core in A520 with Hubble Space Telescope: The Mystery Deepens}",
journal = {The Astrophysical Journal},
}

@ARTICLE{DMok6,
       author = {C. Conroy and Risa H. Wechsler and Andrey V. Kravtsov},
        title = "{Modeling Luminosity-dependent Galaxy Clustering through Cosmic Time}",
      journal = {The Astrophysical Journal},
         year = 2006,
       volume = {647},
       number = {1},
        pages = {201-214},
          }

@BOOK{DMok7,
       author = {Oliver Piattella},
        title = "{Lecture Notes in Cosmology}",
         year = 2018,
         publisher = "Springer International Publisher"
}

@ARTICLE{directdm0,
       author = {{Bertone}, Gianfranco and {Hooper}, Dan},
        title = "{History of dark matter}",
      journal = {Reviews of Modern Physics},
         year = 2018,
       volume = {90},
         }

@ARTICLE{directdm1,
       author = {{Liu}, Jianglai and {Chen}, Xun and {Ji}, Xiangdong},
        title = "{Current status of direct dark matter detection experiments}",
         year = 2017,
         volume = {13},
        journal={Nature Physics},
          }

@article{directdm2,
   title={Dark matter direct-detection experiments},
   volume={43},
   journal={Journal of Physics G: Nuclear and Particle Physics},
   author={Undagoitia, Teresa Marrodán and Rauch, Ludwig},
   year={2015},
}

@article{directdm3,
  title = "{Excess electronic recoil events in XENON1T}",
  author = {Aprile, E. et al.},
  collaboration = {XENON Collaboration},
  journal = {Phys. Rev. D},
  volume = {102},
  year = {2020},
}

@MISC{directdm4,
    author = "J. Aalbers and others",
    title = "{First Dark Matter Search Results from the LUX-ZEPLIN (LZ) Experiment}",
    howpublished = "e-print arXiv:hep-ex: 2207.03764",
    year = "2022",
}

@article{directdm5,
    author = "Viel, Simon",
    collaboration = "DEAP-3600",
    title = "{Dark matter search results from DEAP-3600 at SNOLAB}",
    journal = "PoS",
    volume = "ICHEP2020",
    pages = "655",
    year = "2021",
}

@article{directdm6,
  title = "{Dark Matter Search Results from the PandaX-4T Commissioning Run}",
  author = {Meng, Yue et al.},
  collaboration = {PandaX-4T Collaboration},
  journal = {Phys. Rev. Lett.},
  volume = {127},
  year = {2021},
  }

@article{LCDM1,
year = {2021},
volume = {38},
author = {C Krishnan and R Mohayaee and E O Colgain and M M Sheikh-Jabbari and L Yin},
title = "{Does Hubble tension signal a breakdown in FLRW cosmology?}",
journal = {Class. Quantum Grav.},
}

@article{LCDM2,
title = "{Cosmology intertwined: A review of the particle physics, astrophysics, and cosmology associated with the cosmological tensions and anomalies}",
journal = {Journal of High Energy Astrophysics},
volume = {34},
pages = {49-211},
year = {2022},
author = {Elcio Abdalla and others},
}

@article{LCDM3,
	author = {{Kroupa, P.} and {Famaey, B.} and {de Boer, K. S.} and {Dabringhausen, J.} and {Pawlowski, M. S.} and {Boily, C. M.} and {Jerjen, H.} and {Forbes, D.} and {Hensler, G.} and {Metz, M.}},
	title = "{Local-Group tests of dark-matter concordance
          cosmology - Towards a new paradigm for structure formation}",
	journal = {A\&A},
	year = 2010,
	volume = 523,
}

@article{Gal1,
    author = {Harvey, David and Courbin, F. and Kneib, J. P. and McCarthy, Ian G.},
    title = "{A detection of wobbling brightest cluster galaxies within massive galaxy clusters}",
    journal = {Monthly Notices of the Royal Astronomical Society},
    volume = {472},
    number = {2},
    pages = {1972-1980},
    year = {2017},
}

@article{Gal2,
    author = {Asencio, Elena and Banik, Indranil and Mieske, Steffen and Venhola, Aku and Kroupa, Pavel and Zhao, Hongsheng},
    title = "{The distribution and morphologies of Fornax Cluster dwarf galaxies suggest they lack dark matter}",
    journal = {Monthly Notices of the Royal Astronomical Society},
    volume = {515},
    pages = {2981-3013},
    year = {2022},
}

@article{Gal3,
author = {Oliver Müller  and Marcel S. Pawlowski  and Helmut Jerjen  and Federico Lelli },
title = "{A whirling plane of satellite galaxies around Centaurus A challenges cold dark matter cosmology}",
journal = {Science},
volume = {359},
pages = {534-537},
year = {2018},
}

@article{Gal4,
author = {Michal Bílek and Ingo Thies and Pavel Kroupa and Benoit Famaey},
title = "{Are disks of satellites comprised of tidal dwarf galaxies?}",
journal = {Galaxies},
volume = {9},
pages = {100},
year = {2021}
}

@article{Gal5,
	author = {{M\"uller, Oliver} and others},
	title = "{The dwarf galaxy satellite system of Centaurus A}",
	journal = {A\&A},
	year = 2019,
	pages = "A18",
}

@article{Gal6,
year = {2021},
volume = {917},
number = {2},
pages = {L18},
author = {Sanjaya Paudel and Suk-Jin Yoon and Rory Smith},
title = "{A Corotating Group of Dwarf Galaxies around NGC 2750 as a Centaurus A Analog}",
journal = {The Astrophysical Journal Letters},
}

@article{Gal7,
    author = {Lynden-Bell, D.},
    title = "{Dwarf Galaxies and Globular Clusters in High Velocity Hydrogen Streams}",
    journal = {Monthly Notices of the Royal Astronomical Society},
    volume = {174},
    pages = {695-710},
    year = {1976},
}

@article{Gal8,
	author = {N. Heesters and others},
	title = "{Flattened structures of dwarf satellites around massive host galaxies in the MATLAS low-to-moderate density fields}",
	journal = {A\&A},
	year = 2021,
	volume = 654,
	pages = "A161",
}

@article{JWST1,
year = {2022},
volume = {938},
pages = {L15},
author = {Marco Castellano and others},
title = "{Early Results from GLASS-JWST. III. Galaxy Candidates at z $\sim$9–15$*$}",
journal = {The Astrophysical Journal Letters}
}

@article{JWST2,
year = {2022},
volume = {940},
pages = {L14},
author = {Rohan P. Naidu and others},
title = "{Two Remarkably Luminous Galaxy Candidates at z $\approx$ 10–12 Revealed by JWST}",
journal = {The Astrophysical Journal Letters}
}

@article{JWST3,
year = {2023},
author = {Ivo Labbé and others },
title = "{A population of red candidate massive galaxies $\sim$600 Myr after the Big Bang}",
journal = {Nature}
}

@article{JWST4,
    author = {Lovell, Christopher C and Harrison, Ian and Harikane, Yuichi and Tacchella, Sandro and Wilkins, Stephen M},
    title = "{Extreme value statistics of the halo and stellar mass distributions at high redshift: are JWST results in tension with $\Lambda$CDM?}",
    journal = {Monthly Notices of the Royal Astronomical Society},
    volume = {518},
    number = {2},
    pages = {2511-2520},
    year = {2022},
}

@article{JWST5,
    author = {Michael Boylan-Kolchin},
    title = "{Stress testing $\Lambda$CDM with high-redshift galaxy candidates}",
    journal = {Nature Astronomy},
    year = {2023},
}

@article{JWST6,
    author = {Clara Giménez-Arteaga and others},
    title = "{Spatially Resolved Properties of Galaxies at 5 < z < 9 in the SMACS 0723 JWST ERO Field}",
    journal = {The Astrophysical Journal},
    year = {2023},
    volume = {948}
}

@article{JWST7,
    author = {Amruth, A. and Broadhurst, T. and Lim, J.  and others},
    title = "{Einstein rings modulated by wavelike dark matter from anomalies in gravitationally lensed images}",
    journal = {Nature Astronomy},
    year = {2023},
    volume = {7}
}

@ARTICLE{GCs,
       author = {{Corbelli}, Edvige and {Salucci}, Paolo},
        title = "{The extended rotation curve and the dark matter halo of M33}",
      journal = {Monthly Notices of the Royal Astronomical Society},
         year = {2000},
       volume = {311},
        pages = {441-447},
}

@ARTICLE{Planck,
       author = {{Planck Collaboration} and others},
        title = "{Planck 2018 results. I. Overview and the cosmological legacy of Planck}",
      journal = {A\&AP},
         year = 2020,
       volume = {641},
          }

@MISC{LCDM4,
    author = "P. K. Aluri and others",
    title = "{Is the Observable Universe Consistent with the Cosmological Principle?}",
    howpublished = "e-print arXiv:astro-ph: 2207.05765",
    year = "2023",
}

@article{Costa,
  author = "Costa, L. Filipe O. and Nat\'ario, Jos\'e and Frutos-Alfaro, F. and Soffel, M.",
  title = "Reference frames in General Relativity and the galactic rotation curves",
   journal = {Phys. Rev. D},
  volume = {108},
  issue = {4},
  year = {2023},
}

@BOOK{HawkingEllis,
   author       = {S.W. Hawking and G.F.R. Ellis},
   year         = 1973,
   title        = {The large scale structure of space-time},
   publisher    = {Cambridge Monograph on Mathematical Physics}
}

@ARTICLE{Topology1,
       author = {David B. Malament},
        title = "{The class of continuous timelike curves determines the
                topology of spacetime}",
      journal = {J. Math. Phys.},
         year = 1977,
       volume = {18},
          }

@ARTICLE{Topology2,
       author = {J. L. Friedman and K. Schleich and D. M. Witt},
        title = "{Topological Censorship}",
      journal = {Phys. Rev. Lett.},
         year = 1993,
       volume = {71},
          }

@ARTICLE{Phase1,
       author = {T. W. B. Kibble},
        title = "{Topology of cosmic domains and strings}",
      journal = {J. Phys. A: Math. Gen. A},
         year = 1976,
       volume = {9},
          }

@BOOK{BezzeraBook,
        author = {V. B. Bezerra and E. P. S. Shellard},
        title = "{Cosmic strings and Other Topological Defects}",
        year = 1994,
        publisher    = {Cambridge University Press},
          }

@ARTICLE{TopDefectAtom1,
       author = {G. de A Marques and V. B. Bezerra},
        title = "{Non-relativistic quantum systems on topological defects spacetimes}",
      journal = {Class. Quantum Grav.},
         year = 2002,
       volume = {19},
          }

@ARTICLE{TopDefectAtom2,
       author = {G. de A Marques and V. B. Bezerra},
        title = "{Hydrogen atom in the gravitational fields of topological defects}",
      journal = {Phys. Rev. D},
         year = 2002,
       volume = {66},
          }

@ARTICLE{TopDefectLensing1,
       author = {J. G. de Assis and C. Furtado and V.B. Bezerra},
        title = "{Loop variables, gravitational Aharonov-Bohm effect and gravitomagnetism}",
      journal = {Gravitation and Cosmology},
         year = 2004,
       volume = {10},
          }

@ARTICLE{TopDefectLensing2,
       author = {M. Nouri-Zonoz and A. Parvizi},
        title = "{Gaussian Curvature and Global effects: gravitational Aharonov-Bohm effect revisited}",
      journal = {Phys. Rev. D},
         year = 2013,
       volume = {88},
          }

@ARTICLE{Phase2,
       author = {R. Durrer},
        title = "{Topological defects in cosmology}",
      journal = {New Astronomy Reviews},
         year = 1999,
       volume = {43},
          }

@BOOK{Conical1,
   author       = {C. J. S. Clarke},
   year         = 1994,
   title        = {The Analysis of Space-Time Singularities},
   publisher    = {Cambridge University Press},
}

@ARTICLE{Conical2,
       author = {G. Oliviera-Neto},
        title = "{Identifying Conical Singularities}",
      journal = {J. Math. Phys.},
         year = 1996,
       volume = {37},
          }

@ARTICLE{EllisSchimdt,
   author       = "G. F. R. Ellis and B. G. Schmidt",
   title        = "Singular Space-Times",
   journal      = "General Relativity and Gravitation",
   volume       = "8",
   pages        = "915-953",
   year         = "1977",
}

@BOOK{QM1,
   author       = {C. Cohen-Tannoudji and B. Diu and F. Laloe},
   year         = 1977,
   title        = {Quantum Mechanics, Volume 1},
   publisher    = {Wiley-VCH},
}

@BOOK{QM2,
   author       = {C. Cohen-Tannoudji and B. Diu and F. Laloe},
   year         = 1977,
   title        = {Quantum Mechanics, Volume 2: Angular Momentum, Spin, and Approximation Methods},
   publisher    = {Wiley-VCH},
}

@BOOK{QM3,
   author       = {G. Teschl},
   year         = 2014,
   title        = {Mathematical Methods in Quantum Mechanics},
   publisher    = {American Mathematical Society},
}

@BOOK{MacDonald,
   author       = {F. Bowman},
   year         = 2010,
   title        = {Introduction to Bessel Functions},
   publisher    = {Dover Publications},
}

@ARTICLE{vSB,
   author       = "L. Bratek and J. Jalocha and M. Kutschera",
   title        = "Singular Space-Times",
   journal      = "Phys. Rev. D",
   volume       = "75",
   year         = "2007",
}

@ARTICLE{vSBclass1,
   author       = "W. J. van Stockum",
   title        = "The Gravitational Field of a Distribution of Particles Rotating about an Axis of Symmetry",
   journal      = "Proc. Roy. Soc. Edin.",
   volume       = "57",
   year         = "1937",
}

@ARTICLE{vSBclass2,
   author       = "W. B. Bonnor",
   title        = "A rotating dust cloud in general relativity",
   journal      = "J. Phys. A: Math. Gen.",
   volume       = "10",
   year         = "1977",
}

@ARTICLE{vSBclass3,
   author       = "W. B. Bonnor and B. R. Steadman",
   title        = "Exact solutions of the Einstein-Maxwell equations with closed timelike curves",
   journal      = "Gen. Relativ. Grav.",
   volume       = "37",
   year         = "2005",
}

@ARTICLE{vSBclass4,
   author       = "W. B. Bonnor",
   title        = "Rotating dust clouds in general relativity",
   journal      = "Gen. Relativ. Grav.",
   volume       = "40",
   year         = "2008",
}

@BOOK{Gravitation,
   author       = {C. W. Misner and K. S. Thorne and J. A. Wheeler},
   year         = 1973,
   title        = {Gravitation},
   publisher    = {W.H.Freeman and Co Ltd},
}

@BOOK{Singular0,
   author       = {S. Hawking and R. Penrose},
   year         = 1996,
   title        = {he Nature of Space and Time},
   publisher    = {Princeton University Press},
}

@ARTICLE{Singular1,
   author       = "R. Penrose",
   title        = "Gravitational Collapse and Space-Time Singularities",
   journal      = "Phys. Rev. Lett.",
   volume       = "14",
   year         = "1965",
}

@ARTICLE{Singular2,
   author       = "S. Hawking",
   title        = "Occurrence of Singularities in Open Universes",
   journal      = "Phys. Rev. Lett.",
   volume       = "15",
   year         = "1965",
}

@ARTICLE{Singular3,
   author       = "S. Hawking",
   title        = "Singularities in the Universe",
   journal      = "Phys. Rev. Lett.",
   volume       = "17",
   year         = "1966",
}

@ARTICLE{Singular4,
   author       = "S. Hawking and R. Penrose",
   title        = "The singularities of gravitational collapse and cosmology",
   journal      = "Proc. Roy. Soc. Lond. A.",
   volume       = "314",
   year         = "1970",
}

@ARTICLE{Singular5,
   author       = "F. J. Tipler",
   title        = "Singularities and causality violation",
   journal      = "Annals of Physics",
   volume       = "108",
   year         = "1977",
}

@ARTICLE{Singular6,
   author       = "J. M. M. Senovilla and D. Garfinkle ",
   title        = "The 1965 Penrose singularity theorem",
   journal      = "Class. Quantum Grav.",
   volume       = "32",
   year         = "2015",
}

@ARTICLE{Orientability,
   author       = "C. J. Isham",
   title        = "Spinor Fields in Four Dimensional Space-Time",
   journal      = "Proc. Roy. Soc. Lond. A.",
   volume       = "364",
   year         = "1978",
}

@article{MOND2,
	author = {S. S. McGaugh},
	title = "{The Third Law of Galactic Rotation}",
	journal = {Galaxies},
	year = 2014
}

@article{MOND3,
	author = {S. Trippe},
	title = "{The ‘Missing Mass Problem’ in Astronomy and the Need for a Modified Law of Gravity}",
	journal = {Zeitschrift für Naturforschung A},
	year = 2014
}

@article{MOND4,
	author = {S. S. McGaugh},
	title = "{A tale of two paradigms: the mutual incommensurability of $\Lambda$CDM and MOND}",
	journal = {Canadian Journal of Astrophysics},
	year = 2015
}

@article{MOND5,
	author = {X. Wu and P. Kroupa},
	title = "{Galactic rotation curves, the baryon-to-dark-halo-mass relation and space–time scale invariance}",
	journal = {Monthly Notices of the Royal Astronomical Society},
	year = 2015,
        volume = 446
}

@article{MOND6,
	author = {P. Kroupa},
	title = "{Galaxies as simple dynamical systems: observational data disfavor dark matter and stochastic star formation}",
	journal = {Canadian Journal of Physics},
	year = 2015,
        volume = 93
}

@article{MOND7,
	author = {H. Haghi and A. E. Bazkiaei and A. Hasani Zonoozi and P. Kroupa},
	title = "{Declining rotation curves of galaxies as a test of gravitational theory}",
	journal = {Monthly Notices of the Royal Astronomical Society},
	year = 2016,
        volume = 458
}

@article{MOND8,
	author = {M. Haslbauer and I. Banik and P. Kroupa and N. Wittenburg  and B. Javanmardi},
	title = "{The High Fraction of Thin Disk Galaxies Continues to Challenge $\Lambda$CDM Cosmology}",
	journal = {The Astrophysical Journal},
	year = 2022,
        volume = 925
}

@article{MOND9,
	author = {N. Wittenburg and P. Kroupa and I. Banik and G. Candish and N. Samaras},
	title = "{Hydrodynamical structure formation in Milgromian cosmology}",
	journal = {Canadian Journal of Astrophysics},
	year = 2023,
        volume = 523
}

@unpublished{Beordo2023,
	author = {B., William and Crosta, M. and Lattanzi, M. G. and Re-Fiorentin, P. and Spagna, A.},
	note = {Manuscript submitted for publication},
	title = {Geometry-driven and dark-matter-sustained Milky Way rotation curves with Gaia DR3},
	year = {2023}}

@article{David1,
	author = {D. L. Wiltshire},
	title = "{Cosmic clocks, cosmic variance and cosmic averages}",
	journal = {New J. Phys.},
	year = 2007,
        volume = 9,
        pages = 377
}

@article{David2,
  title = {Cosmological equivalence principle and the weak-field limit},
  author = {Wiltshire, D. L.},
  journal = {Phys. Rev. D},
  volume = {78},
  issue = {8},
  year = {2008}
}

@article{David3,
  title = {From time to timescape - Einstein's unfinished revolution},
  author = {Wiltshire, D. L.},
  journal = {Int. J. Mod. Phys. D},
  volume = {18},
  issue = {14},
  year = {2009},
 pages ={2121-2134}
}

@conference{David4,
    title        = {????????},
    author       = {Wiltshire, D. L.},
    year         = 2012,
    month        = {August-September},
    booktitle    = {Cosmology and Gravitation},
    publisher    = {Cambridge Scientific Publishers},
    address      = {Mangaratiba, Rio de Janeiro, Brazil},
    editor       = {M. Novello and S. E. P. Bergliaffa},
    organization = {BSCG}
}

@article{David5,
  title = {What is dust?—Physical foundations of the averaging problem in cosmology},
  author = {Wiltshire, D. L.},
  journal = {Class. Quantum Grav.},
  volume = {28},
  issue = {16},
  year = {2011},
}

@BOOK{Anderson,
   author       = {M. R. Anderson},
   year         = 2002,
   title        = {The Mathematical Theory of Cosmic Strings},
   Subtitle     = {Cosmic Strings in the Wire Approximation},
   publisher    = {CRC Press},
}

@article{CrostaZamo1,
  title = {General Relativistic Observable for Gravitational Astrometry in the Context of the {{Gaia}} Mission and Beyond},
  author = {Crosta, Mariateresa and Geralico, Andrea and Lattanzi, Mario G. and Vecchiato, Alberto},
  year = {2017},
  journal = {Phys. Rev. D},
  volume = {96},
  number = {10},
  pages = {104030},
}

@incollection{CrostaZamo2,
  title = {Testing the General Relativistic Nature of the {{Milky Way}} Rotation Curve with {{Gaia DR2}}},
  booktitle = {The {{Sixteenth Marcel Grossmann Meeting}}},
  author = {Crosta, Mariateresa},
  year = {2022},
  pages = {3970--3981},
  publisher = {{WORLD SCIENTIFIC}},
}

\end{document}